# EXPLORING THE NUANCES IN THE RELATIONSHIP "CULTURE-STRATEGY" IN THE BUSINESS WORLD

Author: Kiril Dimitrov

*Abstract: The current article explores interesting, significant and recently identified nuances in the relationship "culture-strategy". The shared views of leading scholars at the University of National and World Economy in relation with the essence, direction, structure, role and hierarchy of "culture-strategy" relation are defined as a starting point of the analysis. The research emphasis is directed on recent developments in interpreting the observed realizations of the aforementioned link among the community of international scholars and consultants, publishing in selected electronic scientific databases. In this way a contemporary notion of the nature of "culture-strategy" relationship for the entities from the world of business is outlined.*

*Keywords: Strategic management, strategy, organization culture, entrepreneurship.*

*JEL: L1, M14, L26.*

## INTRODUCTION

A group of three prominent scholars from the University of National and World Economy (UNWE) constitute the main source of supplying knowledge for the potential realizations of the relationship "culture-strategy", arising in (or associated to some extent with) the world of business. As a whole the results of their research work constitute a thorough review of key nuances in the mentioned relationship, already identified by prominent international explorers and/or consultants in respective publications the majority of which date back from the 1970s to the late 1990s of the previous century. It is observed a sporadic use of foreign literature from the first two decades of the 21$^{st}$ century in their literature reviews. This body of scientific literature represents the whole repository in this field, serving as an intersection between cultural studies and strategic management and may be used as a starting point of doing a relevant research in this sphere. The issues, devoted to potential realizations of the relationship "culture-strategy", have been considered as very important for the survival and successful development of business organizations since the times of the energy crisis from the 1970s. But the array of proposed specific solutions and applications within it has been transforming its nature due to changing business environment conditions (globalization, turbulence, terrorism, and climatic change), emerging unique intra-organizational challenges, and accumulation of knowledge and skills in the sphere. That is why the newest achievements in concern of "culture-strategy" relationship apprehensions come of great importance to the scientific community and the managers from the business world. Furthermore, most of the time this relationship occupies a peripheral part of the syllabuses, associated with Business culture and/or Business strategy lecturing at the University. Most of the time it leaves the impression on the unbiased learner or reader of a "magic mantra" for business success in the organization. But these desired results are difficult to be achieved, become evident after a certain (long) period of time, and require taking in great efforts by managers, employees and outside consultants. This represents additional reason of initializing a deliberate study of the recent developments in this field as the appropriate mental programming of the future business and society leaders comes of great importance, concerning the shortage of knowledgeable, skillful and capable professionals in key vocational areas. The planned review of the recent elaborations, concerning the relationship "culture-strategy", is performed through searching of relevant (key) and modern free-access publications (with period of issue: 2011- 06.2016) within the first three electronic pages in accessible electronic scientific databases as Directory of open access journals, Ideas Repec, Proquest, ScienceDirect, Ebsco and Google Scholar. It is detected that many of the selected scientific publications are indexed in more than one specialized database. The association of an article with a database is made by its first encounter within the performed consecutive search in these electronic repositories of appropriate information for the purpose of the current research.

1. **The constitution of the local point of view to the existence of the relationship "culture-strategy"**

The intensive search in the UNWE's library for scientific publications[1] at least partially oriented to the exploration of the relationship "culture-strategy" revealed that just a group of three local colleagues have demonstrated any scientific and/or lecturer's interest in this field since 1990 and up to the moment (07.2016) – the period that marked the transition of Bulgaria to democratic political system, market economy, and country's accession to the European Union. These colleagues are prof. Kiril Todorov, prof. Marin Paunov and prof. Stephen Hristov. Important nuances of their interests in this sphere are outlined through table 1.

---

[1] The performed research was a part of the work tasks, included in a University scientific project, named "Exploring proclaimed firm culture in the virtual realm" (2015-2017).

Table 1. Nuances of demonstrated interest in the relationship "culture-strategy" at UNWE

| Important nuances | Description |
|---|---|
| Identified type of demonstrated interest: | 1. Scientific interest (Paunov, 2015, 2012a, 2005, 1993; Todorov, 2014, 2004, 2000, 1994, 1992; Pivota, Hoy, Todorov, Vojtko, 2011)<br>2. Lecturer's interest (Paunov, 2012b, 1995; Todorov, 2011, 2001, 1997; Hristov, 2009) |
| Identified types of publications: | 1. Article/ collection of articles: (Todorov, 2000, 1992; Pivota, Hoy, Todorov, Vojtko, 2011)<br>2. Monography/ chapter from a book: (Paunov, 2015, 2012a, 2005, 1993; Todorov, 2014, 1994)<br>3. Textbook: (Paunov, 2012b, 1995; Todorov, 2011, 2001, 1997; Hristov, 2009)<br>4. Conference proceedings (Todorov, 2004) |
| Orientation to the discussed field: | 1. Entirely devoted to the explored sphere (Todorov, 2004, 2000)<br>2. The target theme is dwelled on as a chapter from a whole book (Paunov, 2015, 2012a, 2012b, 2005, 1995, 1993; Todorov, 2011, 2001, 1997; Hristov, 2009)<br>3. The subtheme is analyzed to some extent as a part of a larger and wider research (Todorov, 2014, 1994, 1992; Pivota, Hoy, Todorov, Vojtko, 2011)<br>4. Some contents, related in this field may be creatively sought and found even outside the respective chapter (Paunov, 2015, 2012a, 2005; Todorov, 2011, 2001, 1997; Hristov, 2009) |
| Sphere of main interest by which this subtheme is peripherally intersected: | 1. Cultural studies (Paunov, 2015, 2012a, 2005; Todorov, 2000, 1992)<br>2. Strategic management (Hristov, 2009)<br>3. Strategic management of SMEs (Todorov, 2014, 2001, 1997)<br>4. Business strategy (Paunov, 2012b, 1995)<br>5. Entrepreneurship (Todorov, 2011) |
| Predetermined cause-effect direction:<br>(Single chapters from different contributors in a book or collection of articles may provide diverse views to the same issue. That is why some publications are mentioned more than once.) | 1. Culture → strategy (Paunov, 2012b, 1995; Todorov, 1992)<br>2. Strategy → Culture (Todorov, 1992, 1997)<br>3. Both directions are possible (one at a time, contingency approach) – (Paunov 2015, 2012a, 2005; Todorov, 2014, 2011, 2001, 2000, 1992; Hristov, 2009) |
| Applied attributes of professed culture & other important cultural attributes in studying the relationship: | 1. Culture as a phenomenon (Paunov, 2015, 2012a, 2012b, 2005, 1995; Todorov, 2014, 2011, 2000, 2001, 1997, 1992)<br>2. Conventional wisdom (some of its mentioned components may be classified as cultural forms) (Paunov, 2015, 2012a, 2005; Todorov, 2014, 2011; Hristov, 2009)<br>3. Vision (Paunov, 2012b, 1995; Hristov, 2009)<br>4. Mission (Paunov, 2012b, 1995; Todorov, 2001, 1997; Hristov, 2009)<br>5. Memo (Paunov, 2015, 2012a, 2005)<br>6. Corporate philosophy (Paunov, 2012b, 1995)<br>7. A set of cultural forms, i.e. metaphors, myths, jargons, tricks, bluffs, artefact, entrepreneur's personality and behavior (Todorov, 2014, 2011, 2004 2000, 1994)<br>8. Presenting a number of useful definitions for it (Paunov, 2015, 2012a, 2012b, 2005, 1995; Todorov, 2014, 2011, 2000, 2001, 1997, 1992) |
| Identified main topics in exploring the relationship: | 1. Strategy formulation (Todorov, 2001, 2000, 1992; Hristov, 2009)<br>2. Strategy implementation (Todorov, 2001, 2000, 1992)<br>3. Firm performance (Todorov, 2001, 2000, 1992)<br>4. The strategy as a cultural artefact and symbol (Todorov, 2001, 2000, 1992)<br>5. Culture-embedded strategic types of organizations (Todorov, 2001, 2000, 1992; Hristov, 2009)<br>6. Entrepreneurship training (Todorov, 2014, 2011)<br>7. Change management perspective (Paunov, 2012b, 1995; Todorov, 1992; Hristov, 2009) |

Table 1. Nuances of demonstrated interest in the relationship "culture-strategy" at UNWE (continued)

| Important nuances | Description |
|---|---|
| Applied cultural classifications for (in)direct discerning of specific organizational strategic archetypes: (At the early 1990s most of these publications were applied by their disclosure in other author's writings. Later on the respective original publications were found and appropriately cited.) | Harison, 1975 → (Todorov, 1992, 8th article) <br> Handy, 1976, 1985 → (Todorov, 1992, 8th article; 2001) <br> Hebden, 1986 → (Todorov, 1992, 8th article) <br> Miles and Snow, 1978 or Miles, Snow, Meyer, Coleman, 1987 → (Paunov, 2015, 2012a, 2005; Todorov, 1997, 2011, 2001, 2000) <br> Schwartz and Davis, 1981 → (Paunov, 2015, 2012a, 2012b, 2005, 1995; Todorov, 2001, 1997) <br> Quinn and McGrath, 1985 → (Todorov, 2001) <br> Hofstede, 2001 → (Hristov, 2009) <br> Lasher, 2001 → (Hristov, 2009) <br> Hickman and Silva, 1991 → (Hristov, 2009) |
| Applied culture (change) management schemes: | Huse, Cummings, 1985 → (Todorov, 1992, 5th article) <br> Pettigrew, 1990[1] → (Todorov, 1992, 7th article) <br> Baker, 1980 → (Todorov, 1992, 4th article) <br> Sathe, 1983 → (Todorov, 1992, 4th article) <br> Gorman, 1989 → (Todorov, 1992, 1st article) <br> Schein, 1983 → (Paunov, 2012b, 1995) <br> Hickman and Silva, 1991 → (Hristov, 2009) |

Furthermore, a practical interpretation of the hierarchical perspective in analyzing the relationship "culture-strategy" is described in table 2. This perspective is formed by conducting extensive reading and critical analysis of the contents, i.e. presented theories, discussed cases, illustrated examples, etc., from the identified appropriate publications by the aforementioned scholars. The main assumption that is proposed here is that levels of strategy realizations may be presented as attributes, inhabiting a certain cultural level. In this way a rich picture of directly mentioned or implied strategy realizations across cultural levels may be constituted, based on the scientific preferences and assumptions of each of the three scholars. In this way it becomes possible to directly associate a strategy and even some strategic management terms with respective cultural levels and detect the availability of consensus among the scientists in relation with established "cultural level – level of strategy realization or important strategic management terms" associations. The applied set of cultural levels is adapted and evolved in the process of its "matching" with the mentioned strategies and other important strategic management elements[2]. This is how it becomes evident the availability of diverse strategy realizations on predominant number of cultural levels, not only within organizational settings, but also outside them. It deserves mentioning that the global cultural level is missing from the outlined framework and the inclusion of the individual as a cultural unit which is not a frequent practice. It should be drown attention to the fact that organizational and team levels of culture are the richest ones in identified respective strategic elements. The presence of a moderate number of strategic elements is detected at the levels of professional culture, strategic business entity and the individual. Regional and national, and economic sector cultural levels are characterized by lower number of strategic elements as inhabitants, due to the specific research interests, expressed by respective authors.

---

[1] Only basic bibliographical description is provided, but it permits identifying the genuine source in the library (see Pettigrew, 1990).

[2] Detailed critical review of modern theories of cultural levels may be obtained from Dimitrov (2012).

Table 2. The hierarchial perspective in matching culture and strategy

| Observed cultural levels | Corresponding levels of pursued strategy |
|---|---|
| Regional & National levels | 1. Quasi-state development strategies (Hristov, 2009)<br>2. National development strategy (Hristov, 2009)<br>3. Network strategies for business organizations (Hristov, 2009) |
| Culture of an economic sector (industry culture) | 1. Strategic wisdom, strategic thinking, strategic axioms in management (Hristov, 2009)<br>2. Company strategy (Common characteristics of company strategy, possessed by most of the firms, operating in a target industry) (Paunov, 2015, 2012a, 2005) |
| Culture of a professional group/ professional level | 1. Strategic wisdom, strategic thinking, strategic axioms in management (Hristov, 2009)<br>2. Entrepreneur's behavior (Todorov, 2014, 2011)<br>3. Training strategy in the sphere of entrepreneurship (Todorov, 2014, 2011)<br>4. Military strategies (Hristov, 2009) |
| Organizational/ firm corporate level | 1. Strategic attitude to stakeholders (Paunov, 2015, 2012a, 2005)<br>2. Strategic orientation (Hristov, 2009)<br>3. Corporate strategy (Paunov, 2012b, 1995; Todorov, 1992; 2000, Hristov, 2009)<br>4. Business strategy (Paunov, 2012b, 1995; Todorov, 2000, 1992; Hristov, 2009)<br>5. Functional strategy (Hristov, 2009)<br>6. Functional strategy (technology strategy) (Todorov, 1992)<br>7. Functional strategy (culture change management) (Paunov, 2012b, 1995; Todorov, 1992)<br>8. Strategy of SMEs (Todorov, 1992)<br>9. Tactics (Paunov, 2015, 2012a, 2005)<br>10. Activities (Hristov, 2009) |
| Strategic business entity level | 1. Business strategy (Paunov, 2012b, 1995)<br>2. Functional strategy<br>3. Tactics (Paunov, 2015, 2012a, 2005)<br>4. Activities (Hristov, 2009) |
| Team level/ a culture of a business unit/ subculture | 1. Strategic positioning of the company; strategic directions and tasks (Paunov, 2012b, 1995)<br>2. Functional strategy (creativity strategy, quality strategy, production strategy, collaboration strategy) (Hristov, 2005)<br>3. The strategy in SME as a result of the collective mind of the respective entrepreneur and other influential employees (Todorov, 2014, 2011)<br>4. Tactics (Paunov, 2012b, 1995)<br>5. Activities (Hristov, 2009)<br>6. Strategic decision-making (Hritov, 2009) |
| The individual as a cultural unit | 1. Strategic decision-making (Hristov, 2009)<br>2. Entrepreneur's behavior (Todorov, 2014, 2011, 1992)<br>3. Strategic entrepreneurship (Todorov, 2014, 2011)<br>4. Activities (Hristov, 2009) |

## 2. The recent trends of accumulation of scientific results in the field

Nawaser, Shahmehr, Kamel, Vesal (2014) confirm the existence of a significant relationship between strategy and organizational culture. In particular, the research purpose of their publication is to explore the link between organizational culture and Porter's generic competitive strategies (differentiation, low cost and focus) in an Iranian machine-building company. The heading and the abstract give the impression that there is a certain misunderstanding, concerning the direction of the investigated relationship that is evident by comparing the survey result and the aim of study. It is dissolved at the end of the introduction where the respective hierarchy of research questions for the survey is revealed. It becomes evident that the authors support the dual role of organization culture as "double-edged sword", i.e. on one hand, as a facilitating platform in pursuing a company's aims and on the other hand – as change hinder. The performed scientific literature research outlines:

- two nuances in the relationship "culture-strategy", i.e. needed harmony between company culture and its hiring strategy, a positive relationship between the strengthening of a target company culture and an increase in its productivity,
- Porter's generic strategies and value-chain frameworks,
- two useful models are applied for classifying potentially observed organization cultures - Burner and Stalker's Organic-Mechanical Model (see Robbins, 1999)[1] and Robbin's model, consisting in a set of ten lateral interdependent elements of the company as creativity, risk taking, direction, unity and coherence, management support, control, identity, reward system, handling conflict and organizational communication that may applied as a source of diverse structural and behavioral dimensions simultaneously (see Robbins, 1999) (see figure 1).

A specific view to "culture-strategy" relationship is generated by Kibuka-Sebitosi (2015) who localizes it in a larger scheme of complex relationships, i.e. the supermarket industry (supermarket chains) in South Africa and their maintained supply-chain relationships, i.e. the establishment of balance among the interests of the respective constituencies (players) within the respective value chain. In this way an interesting study is designed that has to operate at two cultural levels simultaneously (i.e. organizational level and sector one). Furthermore, culture is just one of the analyzed phenomena – among local context, diversity and institutional dynamics. Identification of best practices for organizational operation in emerging markets in Africa ("cultural hidden secrets and soft skills", related to service management) is pursued not only by theoretical approach as exploration of "culture-strategy" relationship, but also through disclosing the influence of "Ubuntu" philosophy , stakeholder theory and systems thinking. Concerning the observed realizations of the target relationship for this article, it may be concluded that the emphasis is set on the occurrence of the specific relationship "culture-strategy implementation". The performed qualitative analysis is done through the lens of Johnson's culture web model, used as a means of directing author's interest to certain subthemes in concern with the chosen research object, i.e. symbols, organizational structures, rituals and established practices, history and myths, etc.

---

[1] This literature source is missing in the analyzed article. The only one version that is presented here, is Robbins, S. P. (1990). Organization Theory: Structure, Design and Theory (3rd ed.). Prentice-Hall, Englewood Cliffs, NJ.

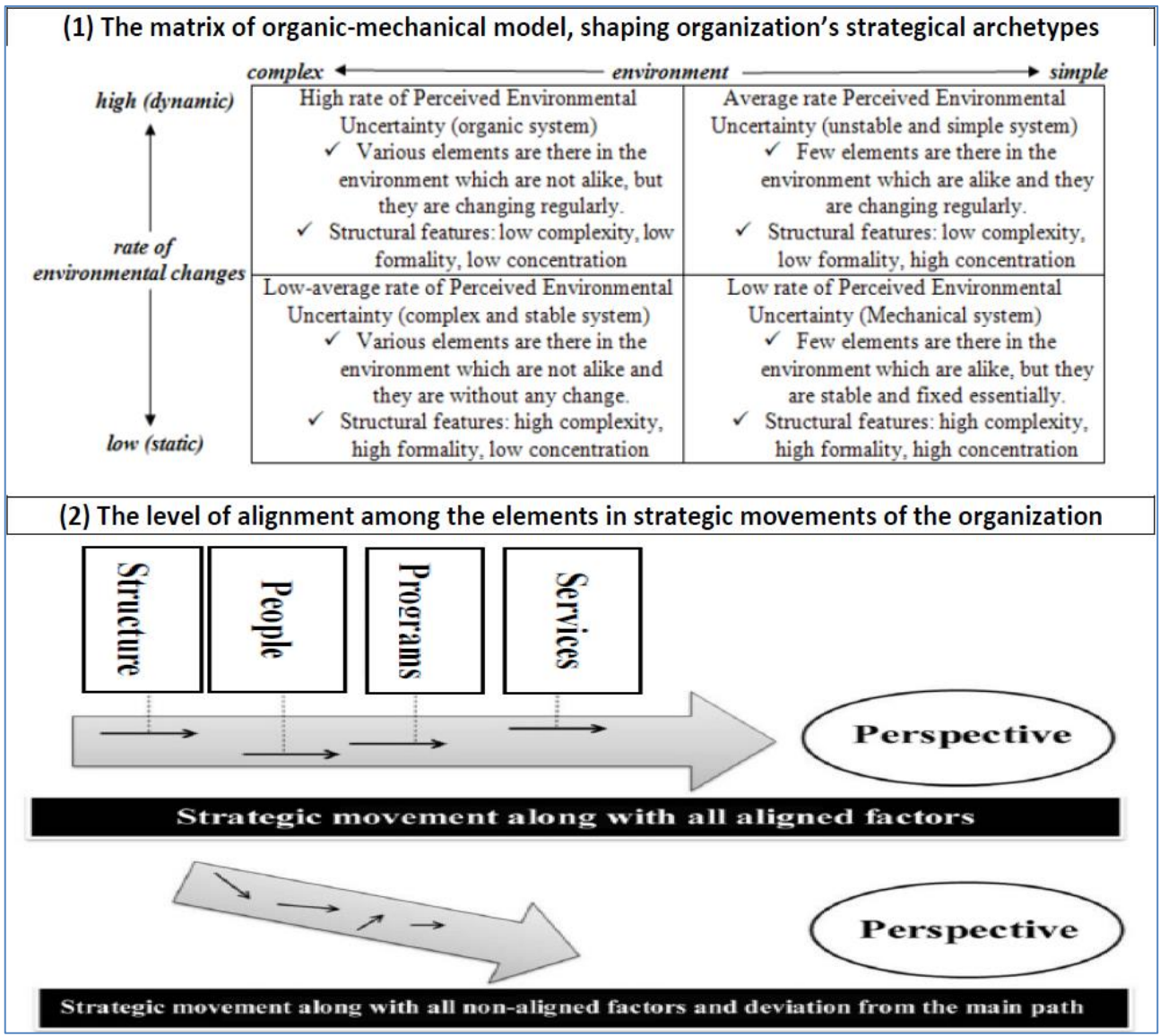

Source: Nawaser, Shahmehr, Kamel, Vesal (2014).

Figure 1. The cultural frameworks, applied for exploring the relationship "culture-strategy"

The need of model's refinement is just mentioned, but not designed, applied and described in the article. There is even no graphical illustration of the applied model. Furthermore, the application of this cultural model is not reviewed as a separate section in the publication and the effectiveness of its use is not mentioned in the conclusion.

The cultural influences at organizational level and economic sector one are directly related to strategic moves in important spheres as building trust, utilizing the potential of cultural diversity, harmonization of daily operational activities with the dominating Ubuntu philosophy, simply surviving competition, defending a leading market position, re-branding, empowering the personnel members, diversification, improving customer service through needs research and added value creation (introducing new products, providing pleasant shopping experience), lowering service costs through increasing the effectiveness of their own operations, managing their supply chains through stimulation of quality improvements among their partners and mitigating the negative effects of inappropriate uses of great bargaining power. In this way strategies, realized at corporate, business and functional level may be identified. The observed competitive strategies may be classified as defending ones and attacking ones. As far as the functional strategies are concerned, the organizational spheres as marketing, human resources and service management may be outlined. The sector level of analysis gives way to the formulation of the collaboration strategy among the supermarket chains and their partners.

A nice mix of theoretical and empirical survey perspectives is incarnated in an article by Cristian-Liviu (2013). The emphasis here is set on the specific nuance of „culture-strategy" relationship, i.e. „dynamic organizational culture – competitive strategy implementation". According to the author this specific culture is deliberately designed in order to sustain undertaken new strategic initiaitves by the senior management of the business organization. Eight organizations from the construction sector in Rumania are defined as a research object. Agroup of 23 people are surveyed through two types of questionnaires, oriented to either identifying the success in completing the strategic intents of the company, or characterizing the existing organizational culture in them. The target firms were chosen to possess a dominating, supportive and dynamic organizational culture with the researcher's aim of checking whether these are more likely to have a strategy formulated and implemented. It seems that this aim to some extent enlarges the target topic, pertaining to strategy.

The cumulative mix of obligatory analysis of a target organization culture and the performing of needed cultural change are presented as the greatest challenge, confronting contemporary managers in their pursuit of desired fit between "new strategy" and "the cultural context present inside the company" Cristian-Liviu (2013). The application of four cultural frameworks is proposed for the solution of the identified complex issue (Morden, 2007, p. 381; Pearce, Robinson, 2007, p. 378-382; Hannagan, 2007, p. 155; McMillan, Tampoe, 2000, p. 234) (see table 3). It has to be marked that Pearce and Robinson (2007) assume the existence of indirect relation between "culture" and "strategy" through key factors, serving as mediating variables that should be managed. They stick to the research tradition that predominated up to the 1970s, assuming culture as a whole, bundled phenomenon.

Table 3. The applied culture analysis frameworks

| Orientation | Framework | Description |
|---|---|---|
| Existing culture analysis | Tony Morden's set of five variables (2007, p. 381)[1] | 1. *Value judgement* - employees' individual and shared judgements, to some extent incarnated in company's mission, objectives and sets of values, direct observed differing behaviours. These values and judgement values define and shape the priorities of the entire strategy formulation and implementation process.<br>2. *Vision*. It represents the means by which the company's judgement values and ideologies are integrated. Its high quality formulation ensures a proper alignment between strategy and organizational culture.<br>3. *Value system* - company's vision and value judgements; the base of defining and implementing the company's mission, strategy and behaviours.<br>4. *Behavioural standards and norms*, i.e. behavior and attitude of employees, directing emerging relations with co-workers, preferred leadership style and supported decision making process.<br>5. *Perceptions regarding needs, priorities and wishes* - influence both the strategic formulation process and the decision making system by establishing the objectives related to sales, market share, quality, value chain management, risk management and ethics. |
| | The relationship between "organizational culture" and the "key factors", influencing the success of the "strategic actions" for achievement of competitive advantage (Pearce, Robinson, 2007, p. 378-382) | The main challenge for managers is the identification of ways in which the existing culture influences the key factors.<br>The list of key factors, influencing organization's core activities and thus – the success of the company, are:<br>1. Structure<br>2. Employees<br>3. Systems<br>4. Managerial style |
| Culture change process | Hannagan's process approach to changing the existing culture (2007, p. 155)[2]. | 1. Unfreezing<br>2. Changing<br>3. Refreezing |
| | The undertaking of changing the culture in order to ensure its total support for the new strategy by means of main activities (McMillan, Tampoe, 2000, p. 234) | The series of main organizational activities:<br>1. Sell the new strategic intent.<br>2. Interpret the existing organizational culture.<br>3. Develop group decision-making skills.<br>4. Introduce innovative mindsets that welcome change.<br>5. Develop skills and knowledge base.<br>6. Encourage staff to feel secure.<br>7. Develop means of helping staff deliver consistent performance.<br>8. Enable accessibility to management during periods of change.<br>9. Encourage thinking that focuses on the outset world. |

Source: Cristian-Liviu (2013).

The empirical perspective confirms the existence of a powerful connection between a dynamic organizational culture and a strategic process intended to gain competitive advantages and that is proved here (see figure 2). It should be noted the undisciplined use of different terms in the sphere of strategic management as strategy, strategic intent and strategic process in pursuing desired competitive advantage ascribes some vagueness to the received results. Here again it can be observed a wide range of strategy realizations – corporate, business and functional ones. Even the scope of functional strategies is a bit enlarged by risk management and ethics.

---

[1] There is no citation for this author in the provided list of references at the end of the article. This information and further literature research proved that this source is identified as Tony Morden's "Principles of strategic management" (see Morden, 2007).

[2] This version of the publication is missing in the references list. The same author is cited by another source - Hannagan, T. (2002), Mastering strategic management, Hampshire: Palgrave. My further literature research identified another source by Tim Hannagan that to great extent may be the original (see Hannagan, 2007).

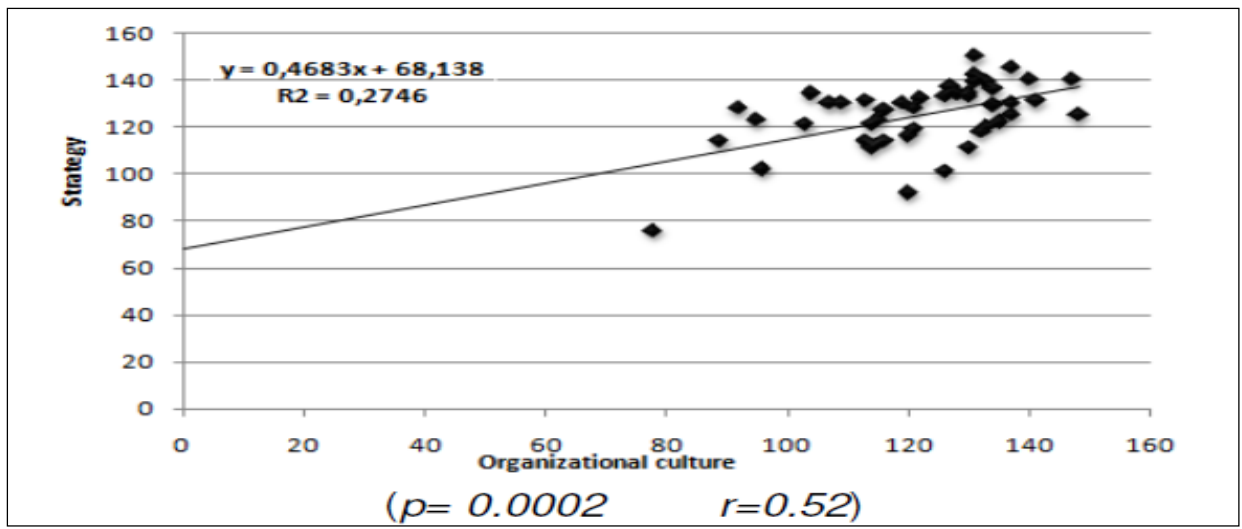

Source: Cristian-Liviu (2013).

Figure 2. The connection between dynamic organizational culture and a strategy, outlined by Pearson's linear correlation coefficient

In his turn, Janićijević (2012) tries to delve into the deep topic of exploring the relationship "culture-strategy" through compiling a review of empirical researches and respective theoretical elaborations, resulting from them. He reaches the conclusion that the aforementioned relationship may be characterized by significant influence, interdependence, mutual conditioning, mutual conformity, compatibility and harmony for the sake of gaining and maintaining a competitive advantage for the business organization (see table 4). The review of representative researches permits Janićijević (2012) classifying recent empiric studies of „culture-strategy" relationship into two groups, grounding the existence of two sub-linkages:

- „general strategies – cultural assumptions and values", and
- „culture - singe functional strategies of an enterprise or strategies within specific business areas, such as human resources management (inducement HRM strategy that includes focus on lowering the costs and involvement HRM strategy that is oriented to innovations and quality), production, marketing (product-marketing strategy) and market entry (innovation strategy, imitation strategy)".

Table 4. Mapping the structure of the relationship „culture-strategy"

| Direction of the influence | Essence of the influence (analyzed spheres, emerging research topics) | Mechanism of the influence | Recommendations to the managers in their pursuit of harmony between the two components in the relationship |
|---|---|---|---|
| *Organizational culture → Strategy* | Strategy formulation process | It shapes the interpretative schemes and meanings which strategic decision-makers assign to the occurrences within and outside of the company, i.e.: (a) It determines the way in which top management gathers information. (b) It determines the way in which they perceive and interpret the environment and the company resources. (c) It influences the way in which they make strategic decisions (strategy selection). | 1. During the process of strategy formulation, managers must take into account dominating cultural assumptions, values and norms in order to provide in advance for the new strategy to comply with them. 2. During the phase of strategic analysis the managers must perform and scan organizational culture profile of the enterprise. 3. During the phase of strategy selection the managers must be ready to adapt the strategy to the existing culture of the company. |
| | Strategy implementation process | It legitimizes or delegitimizes the strategy in sight of managers and employees, depending on the consistency between cultural values and the selected strategy, i.e.: (a) Legitimization - it significantly facilitates strategy implementation. (b) DE legitimization - it makes the implementation of the selected strategy almost impossible. | |
| *Strategy → Organizational culture* | Institutionalization of culture – high levels of conformity with cultural values and norms. | If activities through which the selected strategy is operationalized and implemented are in conformity with cultural values and norms, the strategy will institutionalize and strengthen the existing culture. | 1. The occurrence of a situation in which the managers are forced to select a strategy which is inconsistent with the existing culture, is possible. 2. Managers must be ready and able to close the "cultural gap" during the strategy implementation, i.e. change the existing culture. 3. Managers must develop abilities and knowledge of how to change organizational culture in a planned manner. |
| | Deinstitutionalization of culture – low levels of conformity with cultural values and norms. | Long-lasting and consistent implementation of the selected strategy will deinstitutionalize organizational culture, whereby the process of its change begins. | |

Source: Janićijević (2012).

Several key moments are worth mentioning from the analysis, performed by Janićijević (2012), for the purpose of the current research, as follows:

- The acceptance of mutual conditioning and high correlation between organizational culture and strategy that justifies forbearance from attributing the roles of dependent variable (effect) and independent one (cause) to any of the two counterparts in the target relationship.

- Based on reviewed scientific publications, there may be generated a list of cultural frameworks, applied in the process of exploring the relationship „culture-strategy": (a) adapted Organizational Culture Profile (OCP) instrument (O'Reilly, Chatman, Caldwell, 1991) (six cultural norms: result orientation, detail orientation, support for the people, innovativeness, team orientation and stability); (b) Competing Values Framework (Cameron, Quinn, 1999) (clan culture, hierarchy culture, market culture and adhocracy culture); (c) Wallach classification (Wallach, 1983) (three types of culture: bureaucratic; supportive; innovative or competitive); (d) Hofstede (the set of four dimensions of national cultures analysis: power distance, uncertainty avoidance, individualism–collectivism and masculinity-femininity).

Based on reviewed scientific publications, there may be generated a list of frameworks for strategy operationalization and measurement: (a) Miles and Snow (1978) classification (four types of strategies: prospector, defender, analyzer and reactor); (b) standard marketing strategy classification (three dimensions: differentiation, cost-based efficiency or leading position in costs, and market scope); (c) four competitive goals of the production function - costs, quality, flexibility and reliability, determine chosen production strategies in the company, based on preferred priority of officially stated goals.

Goromonzi, (2016) explores statistically the potential influence of organizational culture and strategy implementation as indices on commercial bank performance, measured by an average of three year period annual return on assets (ROA), in Zimbabwe. The author claims that the presented theoretical model is constructed on the solid ground of the performed literature review, but this is not entirely true, because a cultural framework, outlining attributes as „power, role, task and person", is not described and analyzed there. Furthermore, there are no included publications by Harrison and Handy in the list of references. In fact this model is applied with the purpose of conceptualizing „the intimate link between organizational culture and performance via strategy implementation" (Goromonzi, 2016). More specifically, culture is viewed here as a glue, harmoniously converging „leadership, ethics, strategy and performance". It is assumed that culture realizes a two way interaction with strategic leadership, strategic direction, organizational control and strategy implementation, that contributes to perfection, excellence, integration, problem solving and gradual improvement in the quality of decision-making in the organizations. In this way the performance level of the company as a whole is boosted. The survey confirms that two critical factors – organizational culture and strategy execution, exert a positive and significant impact on performance (figure 3).

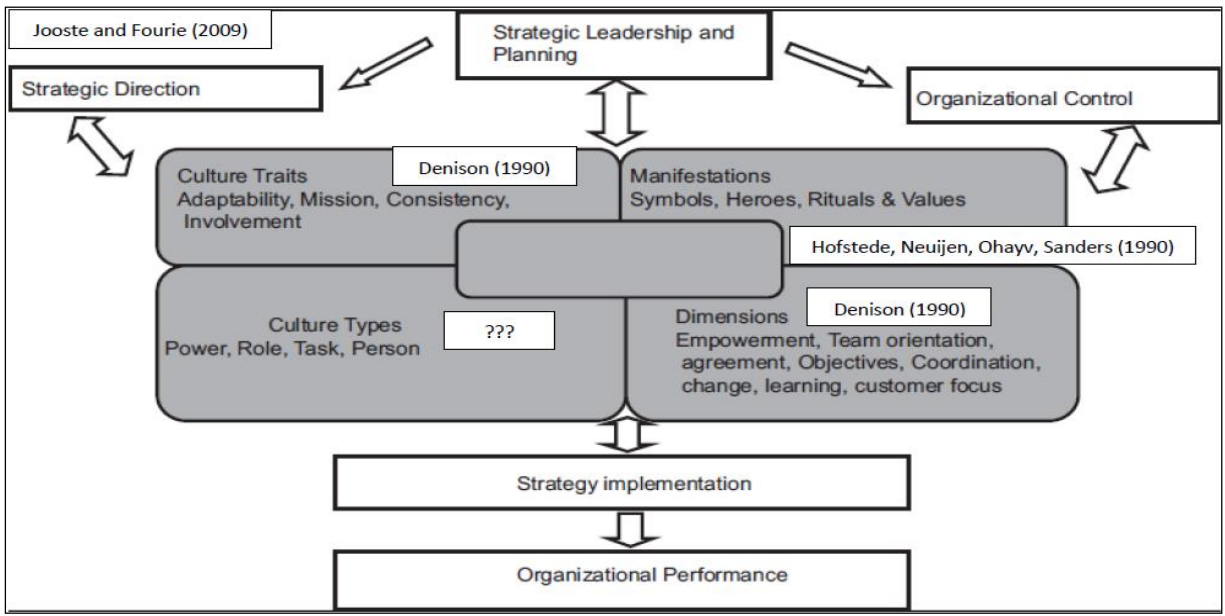

Source: (Goromonzi, 2016).

Figure 4. Goromonzi's interactive dynamics of culture, strategy implementation and performance

The strategic management perspective allows Hanson and Melnyk (2014) to map the full set of alternative opinions, concerning the potential realizations of the relationship "culture-strategy", based on somebody else's or own research results, performed consulting activities and the experience of renowned business organizations, as follows:

- Presenting the stance, expressed by the advocates of failed organization change initiative, i.e. „culture eats strategy for breakfast".
- Assuming culture as "a source of healthy skepticism", characterized as an antidote for "wellknown companies" against potential or pending commercial disasters.
- On the basis of the presented research defending their own stance: the continuous pursuit of „culture-strategy" fit. Such an approach requires from company managers to unlearn specific conventional wisdom attributes in many business-related areas, and basically several management myths (see table 5).

The repeated, double-sided "re-alignment" between "culture" and "strategy", adopted by Hanson and Melnyk (2014) is incarnated in the attributed labor role division between managers, performing within established cultural conditions, and leaders, deliberately struggling to change some of them. The scientists support their approach with the occupied stance of a world famous specialist in the field of cultural studies as Edgar Schein (2010) . An interesting nuance in Hanson and Melnyk's (2014) work is that they have chosen to put an emphasis on organization's efforts in surviving and coping with business environment – paying central attention to the attitudes to the members of the respective value chain, but providing a balanced view to all constituencies as Schein does (Schein, 2010).

Table. 5. „Deceiving" nature of management myths

| Management Myth | The reason(s) it does not work provoke(s) the occurrence of desired results |
|---|---|
| Show them a better way and they will embrace it | Personnel members are not informed properly in order to be able to act efficiently and effectively: (a) they must know how to do the task, and (b) they must know what the purpose of the task (desired outcome) is.<br>Frequently the senior management accept strategic initiatives only as a "how" problem (the same goal; alternative achievement means). That is why employee persuasion and training constitute the preferred array of exerted manageiral influence.<br>Passing the orders down the chain of command leads to emerging of revised goals, set for different functions and departments.<br>This results in selective use of new methods, supporting to great degree the dominating understanding of "what" is to be accomplished.<br>Key obstacles to changing culture:<br>(a) Formulation of a "better" strategy without making efforts for displacement of the existing one.<br>(b) Trying to change team culture without ensuring senior managers' support. |
| Culture and strategy are natural enemies | Combinations of „What?" and „How?":<br>(a) green quadrant of incremental improvement – both elements remain unchanged. HRM methods as training and use of performance-based incentives work well.<br>(b) gray quadrant - one of the key elements must be changed.<br>(c) red quadrant (strategic change) – both elements must be changed.<br><br>**Strategic Initiatives**<br><br>|  | Outcomes Unchanged (What?) | Revised Outcomes |<br>|---|---|---|<br>| Existing Methods | No conflict, incremental changes in efficiency or effectiveness | Uncertainty created, work continues but performance likely sub-optimal, even counterproductive |<br>| Revised Methods (How?) | Conflict created; need for training and reinforcement, not everyone will be able to make transition | Conflict and uncertainty; crisis management required- low likelihood of acceptance of change | |
| Don't tell people how to do their jobs—tell them what you want done and get out of their way | Organization's transition period implies that culture no longer provides the "right" rationale for decision-making.<br>In this case culture should be re-aligned with the changed strategy.<br>Performance measurement and employee rewarding, based simply on tracking achieved results is not suitable, but employee efforts and compliance, and responsibility for the results should become the priority appraisal criteria, applied by managers. |
| The cultural change management efforts should be concentrated within the company | The successful passing through strategic change requires managing the culture change both within the company and in the supply chain. |
| Maintain a restricted number of criteria for selecting partners | *The past*: economics, capacity, and capability considerations.<br>*The current situation increases the number of appraised criteria*: (a) acceptable match of organizational capabilities, (b) acceptable communication levels among the systems, (c) acceptable levels of reading and understanding each other's data, (d) acceptable levels of a cultural match between the respective partners. |
| It does not worth surveying the culture of a potential partner | The historical strengths of a potential partner should be explored in order to grasp its culture.<br>The potential partner's history of success in working with other companies should be checked.<br>Working around a partner's culture requires micro-managing the established relations with detailed contracts, close monitoring and incentives to persuade partner to behave in the right way. |

Source: Hanson and Melnyk (2014).

Ahmadi, Salamzadeh, Daraei, Akbari (2012) focus in exploring the typological and dimensional correlation between organizational culture and strategy implementation, applying Cameron and Quinn's Competing Values Framework model (1999) in a commercial bank in Iran. Based on their research, several nuances in the analyzed relationship may be outlined (see figure 5):

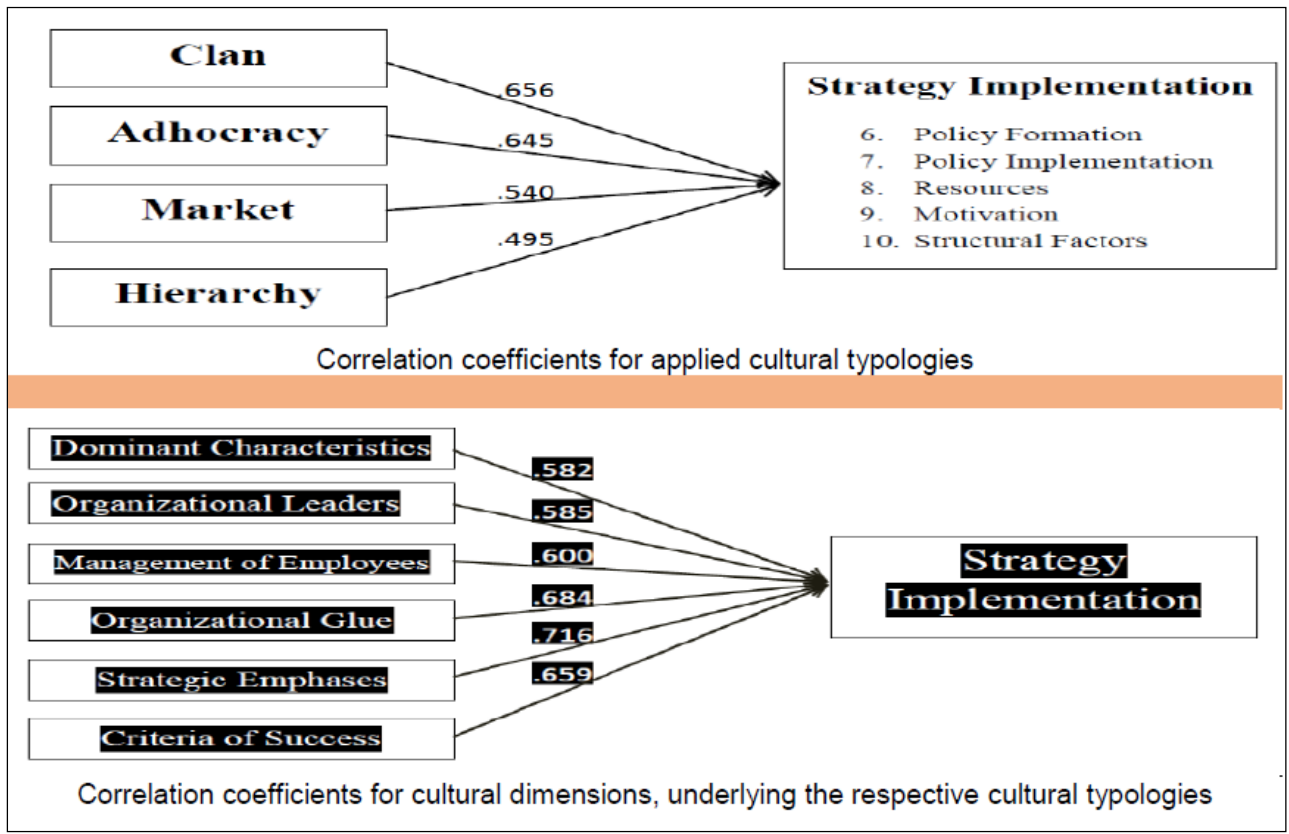

Source: Ahmadi, Salamzadeh, Daraei, Akbari (2012)

Figure 5. The power of the relationships between cultural attributes and strategy implementation

- A meaningful relationship between organizational culture (each one from the four types) and strategy implementation exists. The cultural types, arranged by the decreasing power of their significant relations with implementation process, are: Clan culture, Adhocracy culture, Market culture and Hierarchy Culture. Thus, it becomes evident that flexible types of cultures (Clan, Adhocracy) possess stronger associations with strategy implementation.
- The strategy implementation process is structred into five explanatory dimensions by the authors, as follows: (a) policy formation, (b) policy implementation, (c) resources, (d) motivation, and (e) structural factors.
- All cultural types are correlated to all dimensions of strategy implementation. Furthermore, it is found that the higher degrees of correlations occured when flexible cultures (clan and adhocracy) visited policy formation (dimension 1) and structural factors (dimension 5).

- The organization culture assessment instrument (OCAI), underlying Cameron and Quinn's cultural analysis framework (1999), is applied in this research that leads to paying attention to the respective six cultural dimensions, as follows: (a) dominant characteristics, (b) organizational leaders, (c) management of employees, (d) organizational glue, (e) strategic emphases, and (f) criteria of success.
- It is confirmed that all dimensions of organizational culture are significantly related to strategy implementation. „Strategic emphases" (1st) and „organizational glue" (2d) exercise the greatest effect on the implementation of strategy.

Occupying a strong position that culture has to be aligned with company's strategic intent, Gupta (2011) empirically explores the relationship "organization's strategy – firm culture" as a part of a larger survey, oriented to identifying key characteristics of strategy and culture, and encompassing a group of 32 Indian organizations from economic sectors as construction, banking, information technology, pharmaceuticals, power generation, steel production, and telecommunications. In his endeavors Gupta (2011) relies on:

- Miles and Snow's typology to discern implemented strategy and even grounds his choice by the proven reliability and validity of their approach, its good codification and prediction strengths through other scientist research, and
- Cameron, Quinn and Rohrbaugh's contributions (Cameron, Quinn, 1999; Quinn, 1988; Quinn, Cameron, 1983; Quinn, Rohrbaugh, 1981), forming organizational culture assessment instrument to discern types of dominating firm cultures in separate industries.
- The availability of rich theoretical descriptions that reveal the importance of aligning culture and strategy for organizational performance and effectiveness (Schein, 2004; Sackmann, 1997).

Concerning the scope of the current research, it is found that companies (Gupta, 2011), following prospector strategy are high on adhocracy culture. Both clan and adhocracy cultures were more popular in organizations with analyzer strategy. The firms, pursuing defender and reactor strategy are high on hierarchy and clan culture. Any differences across industries or within a specific economic sector are not provided.

Following Gupta's (2011) perspective to the realization of the relationship "strategy-culture" incites Poore (2015) to put an implicit emphasis on deliberate cultural changes in healthcare organizations. The achievement and sustainable maintenance of the pursued strong "strategy (tactics) – culture" fit is aimed at improving patient experience as a main measure of success in the age of healthcare consumerism. The „ideal organizational culture" for this author may be characterized as „patient driven", „employee-built", „clearly defined", with team members, sharing one vision for „how care is delivered" (Poore, 2015).

Summarizing and critically interpreting Poore's (2015) recommendations for the pursuit of an appropriate fit or unification between culture and strategy leads to constructing an interesting classification of the activities for change agents (table 6).

Table 6. The border between success and failure in pursuing a new strategy in the healthcare sector

| DOs... | DON'Ts... |
|---|---|
| Do weave or hardwire a patient driven culture into the DNA of a target organization (nevertheless a new culture is built or an existing one is redesigned):<br>(a) involve all stakeholders (patients, their families, etc.);<br>(b) appoint employee representatives from all major areas of the patient experience as „architects of cultural blueprints" (clinical and nonclinical, executive and frontline ones). | Do not be just prescriptive for undertaking certain change initiatives. |
| Develop clear end goals, because clarity creates alignment. | Do not launch a one-size-fits-all laundry list of things employees should do or say to improve the patient experience. |
| Empower the constituencies by developing common language and a common set of tools. | Do not assume that culture change is a top-down process. |
| Maintain a consistent decision-making process in regard with the undertaken cultural change, because in this way mutual accountability throughout each step of the patient experience is increased. | Do not define the types of behavior that will no longer be tolerated. |
| Create a culture of "always" and long-term commitment, engaging the entire organization in the process of building something so specific and explicit that all team members know/ see/ feel it. | Do not satisfy yourself with a culture of "sometimes" or shoft-term compliance, concerning the realizations of changes. |
| Build a culture, working horizontally, across all departments, mirroring your patients' experiences with your organization. | Do not overrely on cultural departmental silos. |
| Appropriate conventional wisdom: "This was developed by us and our patients, for us and our patients." | Do not engage many people as architects of new organizational culture. |
| Interpret patient experience as who you are. | Do not interpret patient experience as what you do in the organization. |
| Mantra of the successful cultural transformation: If they build it, they will own it. | Do not accept the pseudo-mantra of the successful cultural transformation as: "If you build it, they will come" |
| Source: Poore (2015). | |

Heracleous, Werres (2016) choose to examine the process that leads to misalignment and eventual corporate failure over time by applying in-depth case studies approach in two American conglomerates. Here, the realization of the relationship „culture-strategy" to some extent takes a certain shape, but indirectly as „patterns" in terms of factors through which misalignments develop, ultimately leading these communication companies to bankruptcy. In their study Heracleous, Werres (2016) apply an elaborated version of "Environment, Strategy, Core Competencies, and Organization" (ESCO) strategic alignment model, by adding "leadership" (including the board of directors) as "the actor guiding alignment, and sustainable advantage as an outcome". The basic term – „strategic alignment", is defined as „the consistency and synergy among the external environment, the strategy, core competencies, and organizational elements such as culture, organization design, processes and people (Heracleous, Wirtz, Pangarkar, 2009). Heracleous, Werres (2016) do not miss to define even the antonym – "strategic misalignment" as "inconsistency or tension among these elements". Concerning the "strategy-culture" relationship it becomes evident that strategy is considered a higher rank construct, while culture is positioned as just one of a group of four organizational elements (see figure 6).

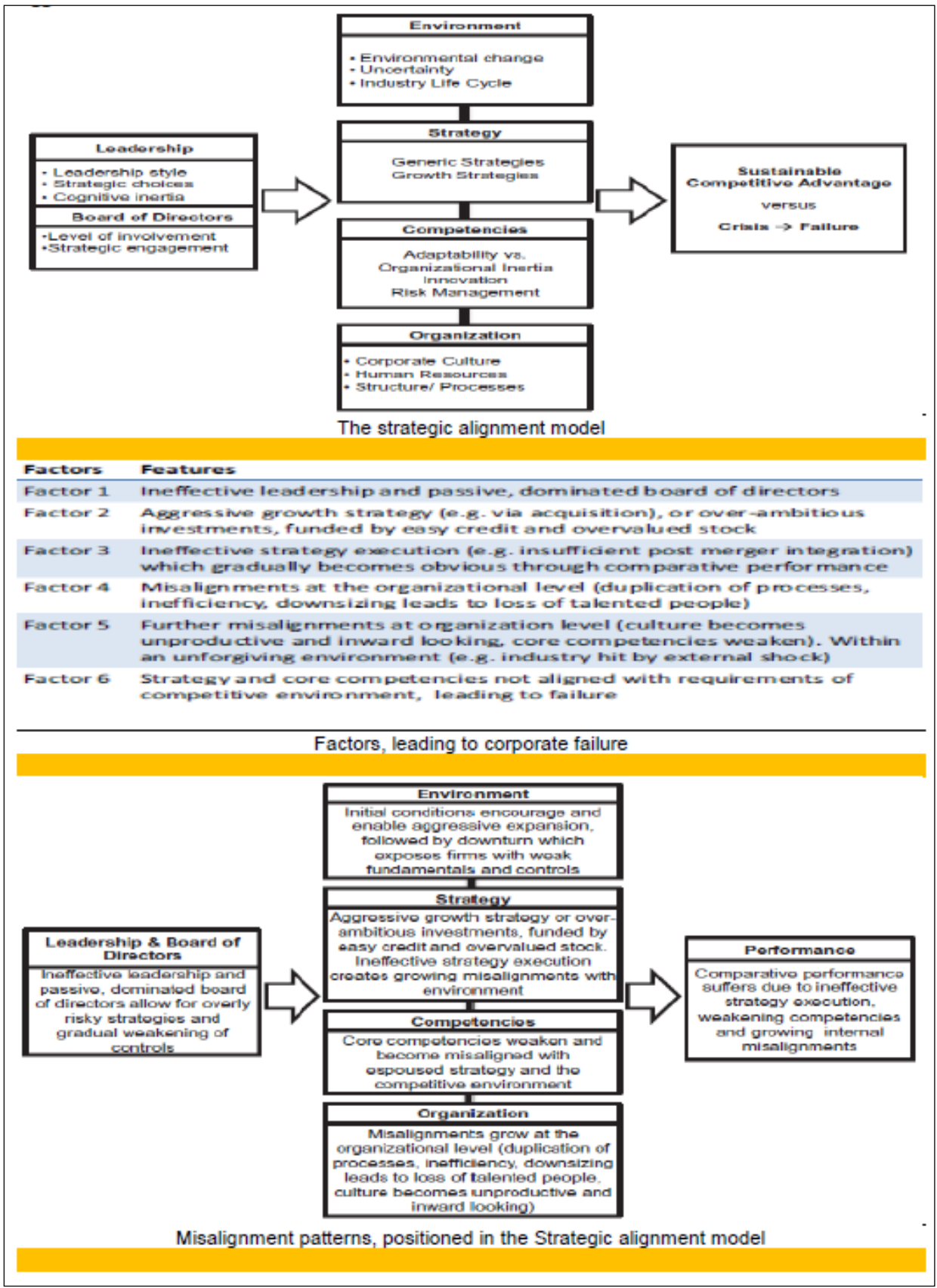

Source: Heracleous, Werres (2016).

Figure 6. The place of culture and strategy among the components of the Strategic alignment model

By utilizing the alignment perspective, the scientists succeed in searching for potential relationships among the aforementioned factors inside the organization setting or outside it within its interaction with the business environment, with or without paying attention to company performance. The element "organization" is indicated to deal with strategy implementation. As a part of it the element "culture" is associated with cultural fit which external orientation (external fit) is considered to address the degree of alignment in a more complicated bundle of bonds, consisting of a larger set of variables as corporate culture, firm's strategy and business environment.

The perspective of business consulting also has a potential to reveal interesting nuances of the "culture-strategy" relationship that is revealed by the experience of two leading business organizations as "Culture Strategy Fit" and "Willis Towers Watson", and a social media post by a renowned senior manager with international experience (Torben Rick, 2013). The first consultancy (Smye, Banack, Lord, 2014; 2014a, 2014b, 2014c, 2014d) puts an emphasis on the most influential factors, contributing to high performance cultures and alignment to strategy in business organizations. Since this array of analytical instruments is "to be bought and sold", their detailed questionnaire design, data collection and retrieval are surrounded with trade secrecy, although company founder's claim that they are based on validated research and measurement of culture at national and organizational level. The applied survey instruments are targeted to global organizations from large to small, and especially to top leaders, strategic HR leaders, innovators, strategy offices and integration teams, and other consulting firms. Concerning the "culture-strategy" relationship, it takes shape through several types of additional strategic focus modules in spheres as customer experience, driving innovation, innovation, operational excellence, performance and results. This predetermines the realization as "culture-functional strategy", aligning dominating cultures at corporate, divisional, organizational, unit and team level with customer relations strategy, innovation strategy, overall company development strategy, production (service) strategy, etc. In fact the newest approach here is represented by the adopted cultural capacities, characterizing the succeeding contemporary organizations (see table 7). On this basis new and interesting (filled in with specific cultural attributes) types of culture may be defined, as agile culture, innovation culture, collaborative culture, etc.

Table 7. Mapping the cultural assessment services, provided by "Culture strategy fit"

| *Cultural patterns (cultural capacities)* | Service: Culture dynamics pulse<br>**Description** | Service: Cultural panoramic snapshot<br>**Description** | Service: Innovation culture pulse<br>**Description** |
|---|---|---|---|
| *Agility* | Adaptability<br>External focus<br>Flexibility<br>Idea generation<br>Learning from experience<br>Risk taking<br>Speed | Adaptability<br>External focus<br>Flexibility<br><br>Learning from experience<br>Risk taking<br>Speed | External focus<br><br><br>Learning from experience<br>Risk taking<br>Speed |
| *Collaboration* | Cooperation<br>Coordination<br>Idea sharing<br>Information sharing<br>Partnering<br>Teamwork | Cooperation<br>Coordination<br><br>Information sharing<br>Partnering<br>Teamwork | Cooperation<br>Coordination<br><br>Information sharing<br>Partnering<br>Teamwork |

Table 7. Mapping the cultural assessment services, provided by "Culture strategy fit" (continued)

| Cultural patterns (cultural capacities) | Service: Culture dynamics pulse<br>**Description** | Service: Cultural panoramic snapshot<br>**Description** | Service: Innovation culture pulse<br>**Description** |
|---|---|---|---|
| *Engagement* | Connection<br>Involvement<br>Mastery<br>Purpose<br>Recognition | Connection<br>Involvement<br>Mastery<br>Purpose<br>Recognition | Connection<br>Involvement<br>Mastery<br>Purpose<br>Recognition |
| *Performance* | Decision-making<br>Excellence<br>Execution<br>Goal alignment<br>Managing performance | Excellence<br><br><br>Managing performance<br>Measuring performance<br>Results focus | Communications<br>Decision-making<br>Excellence<br>Execution<br><br>Managing performance<br><br>Planning |
| *Trust* | Accountability<br>Autonomy<br>Candor<br>Conflict management<br>Fairness & equity | Accountability<br>Autonomy<br>Candor<br>Conflict management<br>Fairness & equity | Accountability<br>Autonomy<br>Candor<br>Conflict management<br>Fairness & equity |
| *Direction* | | Communication<br>Goal alignment<br>Priority setting | |
| *Discipline* | | Compliance<br>Decision making<br>Execution<br>Planning<br>Process discipline | |
| *Risk* | | Managing risk<br>Sustainability | |
| *Innovation* | | | Alignment<br>Idea generation<br>Idea sharing<br>Idea assessment<br>Commercialization<br>Customer focus |
| Sources: (Smye, Banack, Lord, 2014; 2014a, 2014b, 2014c, 2014d). | | | |

The second consulting company - Willis Towers Watson (***, 2016), deliberately concentrates its efforts on identifying the potential relationships between selected cultural aspects and strategic goals of the firm and the appropriate means to further a necessary long-term culture change (i.e. underlying systems, processes, behaviors) in order to establish and maintain sustainability in the levels of employee engagement among its client organizations. By theoretical and empirical research the consultancy tried to cross-match its client organizations' strategic focuses (summarizing how the companies compete, i.e. formulating a set of five core strategies: efficiency strategy, quality strategy, innovation strategy, customer service strategy and reputation/brand strategy), employee survey results (from a period of several decades), measures of company financial performance and thus identified specific, appropriate cultural profiles, contributing in the best way to the achievement of organizational success within each one of the aforementioned strategies (***, 2016). The set of inferred core strategies is formulated in a way that allows simultaneous pursuing of more than one strategy by a certain client organization. In fact, theoretically these "core strategies" may be classified as functional strategies (see figure 7).

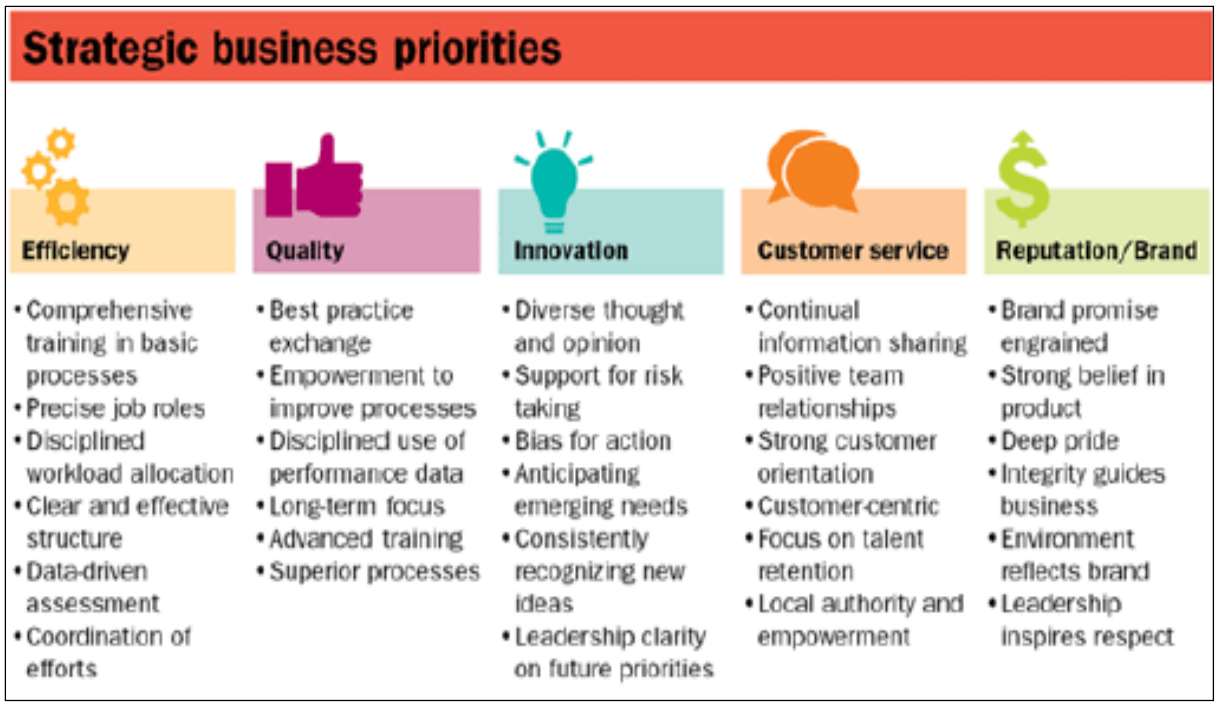

Source: (\*\*\*, 2016).

Figure 7. Specific sets of cultural characteristics, supporting each one of the five core strategies

Keeping the commercial secrets, the consultancy does not miss to elucidate that they provide to their customers the opportunity of discovering their management gaps or achieved degree of strategy/culture alignment, by measuring how closely their current cultures and desired "ideal" future states, align with each of the aforementioned cultural profiles. The crafting or "strategic design" of specific organization's broader talent vision and strategy are defined as the only effective means of influencing a target company culture, based on extensive opinion survey among experts and results from empirical analysis (see figure 8). The creativity in making such a choice of organization change enabler is evident, although the pending issues in concern with operationalization of the term „talent management" still remain (Thunnissen, Boselie, Fruytier, 2013; Scullion, Collings, 2011; Iles, Preece, Chuai, 2010).

In his LinkedIn post Torben Rick (2013) defends the significance of both company strategy and its culture for the acquisition and maintenance of organizational success. He describes key nuances in the „culture-strategy" link by synchronized and aligned cascading of the respective phases in their realizations (figure 9). He leaves the impression that there is even a correspondence between certain implementation phases of both constructs (for example: goals-values; objectives-practices; activities-behaviors) until the desired results are gained. Company's deliberately created and preliminary thought over and debated official mission, vision and values declarations seem to be assigned an initial position to potential realizations of both strategy and culture.

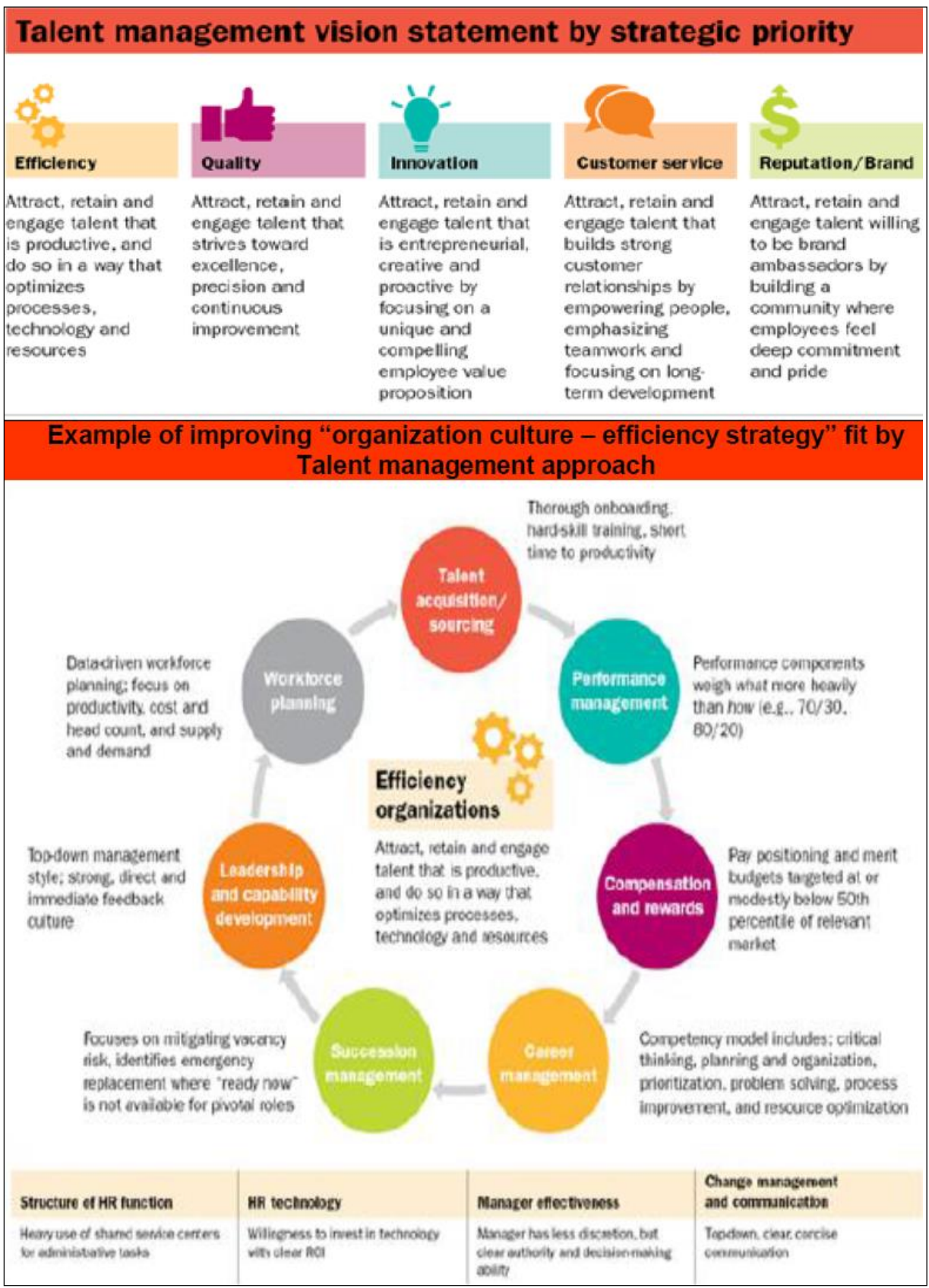

Source: (\*\*\*, 2016).

Figure 8. Talent strategy as an enabler of „firm culture – functional strategy" fit

So, the professed culture of the firm is what is meant here by the construct of "culture", being aware that official documents as declaration of mission, vision and values are in their turn products of the underlying assumptions that dominate among the leaders and the employees in the business organization, i.e. the hidden part of the culture.

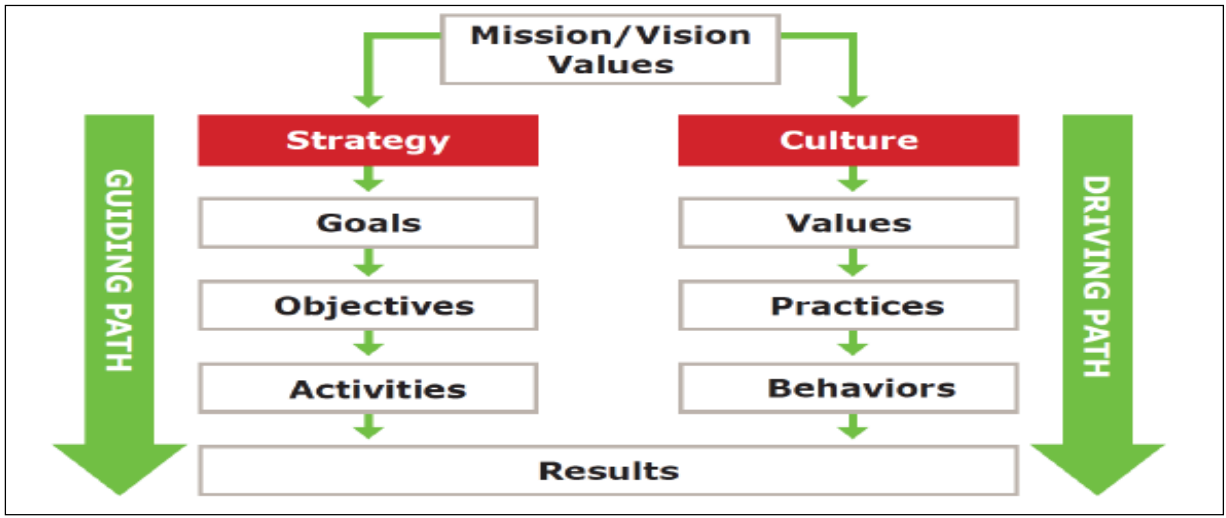

Source: Rick (2013).

Figure. 9 Synchronized and aligned realizations of culture and strategy

The identified lines of relations between culture and strategy, resulting probably to a great extent from Torben Rick's rich experience in the business world, are based on comparing specific characteristics of (and/or differences between) both objects (phenomena, processes, constructs) (see table 8).

Table 8. The array of relationship lines between culture and strategy

| STRTEGY | CULTURE |
|---|---|
| Rational, logical, clear and simple. | It means different things to different people (rich in nuances of meaning). Emotional, ever-changing, complex, human, vulnerable, moody as the people who define it. |
| It should be easy to comprehend and to talk about. | It can be intimidating and frustrating; culture is often subordinated, misunderstood, or unappropriated. |
| Strategy drives focus and direction. | Culture is the emotional, organic habitat in which a company's strategy lives or dies |
| Strategy is just the headline on the company's story. | Culture needs a clearly understood common language to embrace and tell the story that includes mission, vision, values, and clear expectations. |
| Strategy is about intent and ingenuity. | Culture determines and measures desire, engagement, and strategy execution. |
| Strategy lays down the rules for playing the game. | Culture fuels the spirit for how the game will be played. |
| Strategy is imperative for differentiation. | A vibrant culture delivers the strategic advantage. |
| It may be changed frequently, but not periodically or every day. | Culture is built or eroded every day. How you climb the hill and whether it's painful, fun, positive, or negative defines the journey. |
| High degree "culture-strategy" fit. | When culture embraces strategy, execution is scalable, repeatable, and sustainable. |
| A clear competitive advantage. | Formation of an appropriate culture. |

Table 8. The array of relationship lines between culture and strategy (continued)

| STRTEGY | CULTURE |
|---|---|
| Understanding the health and engagement of your organization with pursued strategy | Monitoring of culture |
| A doomed strategy. | A strategy that is at odds with a company's culture. |
| …strategy, change management, innovation, operational efficiency, lean process, vision and mission. | Culture is eating what it kills… |
| Source: Rick (2013). | |

Yarbrough, Morgan, Vorhies (2011) look for intersections among organization theory (especially configuration theory), competing values theory, and fit assessment methodologies in order to explore the existence of product market strategy–organization culture fit and its impact on organization's performance. The successfully performed interpretation of the fit assessment methodologies (Venkatraman, Prescott, 1990) in order to illustrate realizations of "product market strategy – organization culture" relation is especially important to the current research (see table 8).

Table 8. Description of the applied fit methodologies

| Types of fit | Description | Statistical method for data retrieval | Advantages of the statistical method | Disadvantages of the statistical method |
|---|---|---|---|---|
| Fit as covariation | The fit between product market strategy and organizational culture is conceptualized as a second order factor that represents co-alignment between the three first-order factors comprising product market strategy (aforementioned decisions) and the four first-order factors comprising organizational culture (cultural archetypes). | Structural equation model (SEM) | It allows: (a) Identification of a second-order fit factor. (b) Modelling of measurement error. (c) the researcher to examine the relationship between any second-order fit factor identified and business performance dependent variables. | It limits: (a) the researchers' ability to introduce multiple control variables. (b) the use of non-continuous variables. |
| Fit as profile deviation | The degree to which the product market strategy and organizational culture characteristics of a business differ from those of an ideal profile in which they fit together in ways that produce superior performance. Finding ideal profiles requires identifying high performing businesses pursuing different product market strategies, calibrating their organizational culture characteristics as an ideal profile, and assessing strategy-culture fit as deviation from this ideal profile. | Regression analyses | It provides: (a) greater insights into the precise form of the strategy-culture fit relationship. (b) much greater flexibility in the use of multiple control variables. | It provides a lower ability to control for measurement error in comparison to SEM approaches. |
| Source: Yarbrough, Morgan, Vorhies (2011). | | | | |

In fact the relationship among three important elements of a firm's product market strategy (i.e. respective managerial decisions about offering superior product/service, lowering delivered cost, and establishing product-market scope) and four organization culture types (Hierarchy, Adhocracy, Clan, and Market – based on publications as Quinn and Cameron, 1983; Cameron, Freeman, 1991) is examined here. In this way a proposed structure of "culture-strategy" relationship emerges for companies, belonging to US trucking industry, defined as a research object. The existence of interrelationships among product market strategy decisions and organizational culture orientations consistent with configuration theory conceptualizations of product market strategy – organizational culture fit may be defined as a novelty in this research that is proven empirically. Furthermore, significant relationships between product market strategy–organizational culture fit and firms' customer satisfaction and cash-flow return on assets performance are also identified.

**Conclusion**

The already presented analysis of key, recent scientific publications from selected electronic academic databases permits making several conclusions regarding the realizations of "culture-strategy" relationship:

- The topic still attracts scholars' interest as a main subject of research or as an indispensable part of a larger and wider research.
- The concept of predetermined cause-effect direction between the elements of the aforementioned relationship (one-way, two-way) is not rejected, but a greater diversity of the realized interactions is noted by giving reasons for its characterizing as significantly influential, interdependent, mutual conditioning, mutual conforming, compatible and harmonious.
- The portfolio of characteristics for the successful organization culture is not limited only to strong and flexible, but also it encompasses characteristics as dynamic, agile, innovative and collaborative.
- Mission and vision continue to be the chosen attributes of professed culture in studying this relationship.
- The scope of topics in exploring the relationship is a bit enlarged through assuming the aim of „culture-strategy" fit not only as performing a one-time act, but also as a matter of continuous pursuit by managers. The search for potential intersections among organization theory, different cultural theories, and fit assessment methodologies (fit as covariation and fit as profile deviation) show a potential to expand the research realm for the aforementioned relationship and represent a prerequisite for the emerging of new topics.
- The interest to study of "culture-strategy" relationship penetrated in many spheres that is evident by the great diversity of research objects – the business organization as a holistic construct, the international (global) companies, the supermarket industry, its supply-chain, the companies from the banking sector, the IT sector, the

- telecommunications, the construction industry, the machine-building sector, the US tracking industry, the healthcare industry, the pharmaceuticals, the power generation, the steel production, etc.
- The list of applied cultural classifications for (in)direct discerning of specific organizational strategic archetypes includes: Burner and Stalker's Organic-Mechanical Model and Robbins' model (see Nawaser, Shahmehr, Kamel, Vesal, 2014); Johnson's culture web model (see Kibuka-Sebitosi, 2015); Tony Morden's set of variables, Pearce and Robinson's model, relating "organizational culture" and "key factors" (see Cristian-Liviu, 2013); adapted Organizational Culture Profile (OCP) instrument by O'Reilly, Chatman and Caldwell, Competing Values Framework by Cameron and Quinn, Wallach's culture classification, Hofstede's set of four national culture dimensions (see Janićijević, 2012); Cameron and Quinn's Competing Values Framework model, the six dimensions of organization culture assessment instrument (OCAI) (see Ahmadi, Salamzadeh, Daraei, Akbari, 2012); Cameron, Quinn and Rohrbaugh's contributions, Schein's and Sackmann's theoretical descriptions, revealing the importance of aligning culture and strategy for organizational performance and effectiveness (Gupta, 2011); "Environment, Strategy, Core Competencies, and Organization" (ESCO) strategic alignment model (Heracleous, Werres, 2016); shared contributions of Quinn, Cameron and Freeman (Yarbrough, Morgan, Vorhies, 2011).
- The list of applied strategy classifications includes: Porter's generic competitive strategies (differentiation, low cost and focus) (see Nawaser, Shahmehr, Kamel, Vesal, 2014); Miles and Snow's classification (Gupta, 2011; Janićijević, 2012); standard marketing strategy classification, four competitive goals of the production function (see Janićijević, 2012); five explanatory dimensions of strategic implementation process (see Ahmadi, Salamzadeh, Daraei, Akbari, 2012).
- The list of culture change schemes encompasses: Hannagan's process approach to changing the existing culture, McMillan and Tampoe's framework of changing the culture (see Cristian-Liviu, 2013);
- The scope of aims of maintaining the "culture-strategy" fit for the business organization is not limited only to high performance, efficiency, high quality of decision-making processes and productivity, but also includes the gaining and maintaining of a competitive advantage, effectiveness, realization of strategic intents, stakeholders' perspective consideration, conformity with dominating Ubuntu philosophy in Africa, achieving strategic alignment, customer experience, driving innovation and operational excellence, successful post mergers and acquisitions management.
- The predominant level of analysis are: (a) the organizational one and (b) in rare cases the situation in a certain industry is analyzed as a whole.
- A bunch of emerging relationships between different levels of culture and strategy is formed through analyzing the selected group of publications (table 9). It enriches the presented as a starting point array of nuances in the relationship "culture-strategy" on organizational level and in concern with combinations of cultural attributes and strategic elements/ processes.

Table 9. Cultural levels and/or attributes, corresponding to certain levels of strategy

| |
|---|
| *The industry level:* <br> Industry culture – widespread strategy preferences among the entities in an industry (Kibuka-Sebitosi, 2015) |
| *The organization level:* <br> Organization culture – company strategy (Nawaser, Shahmehr, Kamel, Vesal, 2014; Kibuka-Sebitosi, 2015; Hanson, Melnyk, 2014; Smye, Banack, Lord, 2014; 2014a, 2014b, 2014c, 2014d) <br> general strategies – cultural assumptions and values (Janićijević, 2012) |
| *Peculiar combinations:* <br> dynamic organizational culture – competitive strategy implementation (Cristian-Liviu, 2013) <br> Professed culture – strategy (Torben Rick, 2013) <br> organization culture – strategy implementation – performance (Goromonzi, 2016) |
| *Culture versus singe functional strategies of an enterprise or strategies within specific business sub-areas:* <br> company culture and its hiring strategy (Nawaser, Shahmehr, Kamel, Vesal, 2014) <br> culture - human resources management (inducement HRM strategy that includes focus on lowering the costs and involvement HRM strategy that is oriented to innovations and quality) (Janićijević, 2012) <br> culture - production (Janićijević, 2012) <br> culture - marketing (product-marketing strategy) (Janićijević, 2012) <br> "product market strategy – organization culture" (Yarbrough, Morgan, Vorhies, 2011) <br> culture - market entry (innovation strategy, imitation strategy) (Janićijević, 2012) <br> culture - customer relations strategy (Smye, Banack, Lord, 2014; 2014a, 2014b, 2014c, 2014d) <br> culture - innovation strategy (Smye, Banack, Lord, 2014; 2014a, 2014b, 2014c, 2014d) <br> culture - production (service) strategy, etc. (Smye, Banack, Lord, 2014; 2014a, 2014b, 2014c, 2014d) |
| *Cultural attributes versus strategic elements or processes:* <br> Strategic emphases – implementation of strategy (Ahmadi, Salamzadeh, Daraei, Akbari, 2012) <br> „organizational glue" - implementation of strategy (Ahmadi, Salamzadeh, Daraei, Akbari, 2012) <br> Clan culture - implementation of strategy (Ahmadi, Salamzadeh, Daraei, Akbari, 2012) <br> Adhocracy culture implementation of strategy (Ahmadi, Salamzadeh, Daraei, Akbari, 2012) <br> prospector strategy - adhocracy culture (Gupta, 2011) <br> clan and adhocracy cultures - analyzer strategy (Gupta, 2011) <br> defender and reactor strategy - hierarchy and clan culture (Gupta, 2011) <br> "strategy (tactics) – culture" (Poore, 2015) <br> cultural aspects - efficiency strategy (\*\*\*, 2016), <br> cultural aspects - quality strategy (\*\*\*, 2016), <br> cultural aspects - innovation strategy (\*\*\*, 2016), <br> cultural aspects - customer service strategy (\*\*\*, 2016), <br> cultural aspects - reputation/brand strategy (\*\*\*, 2016), |

**Information about the author**

associate professor Kiril Dimitrov, Ph.D., "Industrial business" department, University of National and World Economy (UNWE) – Sofia, Bulgaria, e-mail: kscience@unwe.eu.